\documentclass[preprint]{elsart}

\usepackage{graphicx}

\begin{document}
\begin{frontmatter}

\title{Effects of radiation damage caused by proton irradiation 
on Multi-Pixel Photon Counters (MPPCs)}
\author[NDA]{T.Matsumura\corauthref{cor1}},
\corauth[cor1]{Corresponding author. Tel.: +81-46-841-3810; fax: +81-46-844-5912.}
\ead{toru@nda.ac.jp}
\author[Tech]{T.Matsubara},
\author[Kyoto]{T.Hiraiwa},
\author[Osaka]{K.Horie},
\author[Tech]{M.Kuze},
\author[Nara]{K.Miyabayashi},
\author[Kyoto]{A.Okamura},
\author[RCNP]{T.Sawada}
\author[Osaka]{S.Shimizu},
\author[NDA]{T.Shinkawa},
\author[Kyoto]{T.Tsunemi},
\author[RCNP]{M.Yosoi}

\address[NDA]{
\it Department of Applied Physics, National Defense Academy,
Yokosuka~239-8686, Japan}
\address[Tech]{
\it Department of Physics, Tokyo Institute of Technology, 
Tokyo~152-8551, Japan}
\address[Kyoto]{
\it Department of Physics, Kyoto University, 
Kyoto~606-8502, Japan}
\address[Osaka]{
\it Department of Physics, Osaka University, 
Toyonaka~560-0043, Japan}
\address[Nara]{
\it Department of Physics, Nara Women's University, 
Nara~630-8506, Japan}
\address[RCNP]{
\it Research Center for Nuclear Physics, Osaka University, 
Ibaraki~567-0047, Japan}

\begin{abstract}
We have investigated the effects caused by 
proton-induced radiation damage on Multi-Pixel Photon Counter (MPPC),   
a pixelized photon detector developed by Hamamatsu Photonics. 
The leakage current of irradiated MPPC samples 
linearly increases with total irradiated doses due to radiation damage, 
which is not completely recovered even 
after a year from the irradiation. 
No significant change has been observed in the gains at least up to 8.0~Gy 
(9.1$\times 10^7$~n/mm$^2$ in 1~MeV neutron equivalent fluence, $\Phi_{\rm eq}$). 
The device has completely lost its photon-counting capability  
due to baseline fluctuations and noise pile-up 
after 21~Gy irradiation 
($2.4\times 10^8$~n/mm$^2$ in $\Phi_{\rm eq}$),  
which might be problematic for some applications, such as ring-imaging 
Cherenkov detectors. 
We have found that the pulse-height resolution 
has been slightly deteriorated  
after 42~Gy irradiation (4.8$\times 10^8$~n/mm$^2$ in $\Phi_{\rm eq}$), 
where the measured sample has been illuminated with 
a few hundred photons. 
This effect should be considered 
in the case of energy-measurement applications. 
\end{abstract}


\end{frontmatter}


\section{Introduction}

Pixelized Photon Detector (PPD),
a generic name of multi-pixel Geiger-mode avalanche photodiodes,
is a high-sensitivity photon detector developed recently.
This device has attracted much attention 
since a PPD has great advantages over photo-multipliers, 
such as insensitivity to magnetic fields, compactness, 
low bias voltage and low power consumption, 
although the active area of the device  
is limited to be $1\sim 10$ mm$^2$ at present.
Therefore a PPD device is very useful for  
high-energy and astrophysics experiments and 
medical science applications, 
especially for a readout device with scintillating fibers, 
wave length shifting fibers,
and fine granulated scintillating crystals, etc.
A review about PPD has been  
prepared by D.~Renker~\cite{Renker:2006ay}, for example.

The Multi-Pixel Photon Counter (MPPC) is one 
of such PPD devices which has been developed by Hamamatsu Photonics~\cite{HPK}.  
Three types of MPPCs (100, 400, and 1600-pixels type) 
with an active area of 1$\times$1 mm$^2$, and 
some other larger devices, are commercially 
available at present.
Experimental groups, such as T2K, ILC, Belle, TREK and 
KOTO, are planning to adopt the MPPC in their detector systems 
for scintillating fibers and/or wavelength-shifting fibers readout, 
and for ring imaging 
Cherenkov detectors~\cite{Itow:2001ee, Behnke:2007gj, 
Hashimoto:2004sm, Imazato, Yamanaka}. 
Experimental studies of the basic performance of the MPPC have been 
carried out by those groups (see e.g.~\cite{Gomi:2007zz}).

For practical use of the MPPC, 
radiation hardness is one of the important issues 
to be made clear for applications in a high radiation environment. 
In general, an indication of radiation damage on  
silicon devices is an increase of the leakage current, 
which is caused by lattice defects 
created due to radiation damage.
An experimental study with a positron beam has been reported 
for radiation hardness of old types of MPPCs,
showing an increase of the leakage current 
and the dark-count rates but no significant change 
in the gain and the photon-detection efficiency~\cite{Musienko:2007zz}.
For $\gamma$-ray irradiation, high dark pulses generated at the outer regions 
of individual pixels have been observed after 240 Gy irradiation   
in addition to increases of the leakage current 
and the noise rates~\cite{Matsubara}.
To investigate hadron-induced radiation damage, 
which is important in high-energy physics experiments, 
we performed an irradiation experiment with a 53.3 MeV proton beam. 

In the following part of this report, at first, 
we describe the setup and the procedure for the irradiation experiment. 
Then, characteristic changes, such as the leakage current, 
gains, dark counts, pulse-height distributions, are presented. 
Finally, we discuss issues to be concerned 
with the effects caused by 
proton-induced radiation damage from a practical standpoint.

\section{Experiment}

\subsection{MPPC samples}
Two pieces of MPPC samples, Hamamatsu S10362-11-050C delivered 
in February 2007, were irradiated with a proton beam in the experiment. 
We refer to these samples as ``Sample~\#1'' and ``Sample~\#2''.
The samples have an active area of $1\times 1$ mm$^2$ and 
consist of 400 pixels of Geiger-mode avalanche photodiodes whose
individual pixel-size is 50$\times$50~$\mu$m$^2$.  
The operating voltage, which is the voltage giving 
a gain of 7.5$\times$10$^5$ at 25~$^\circ$C, 
is 69.75~V for Sample~\#1, and 69.58~V for Sample~\#2, 
respectively.

\begin{figure}[t]
  \begin{center}
    \includegraphics[width=7cm]{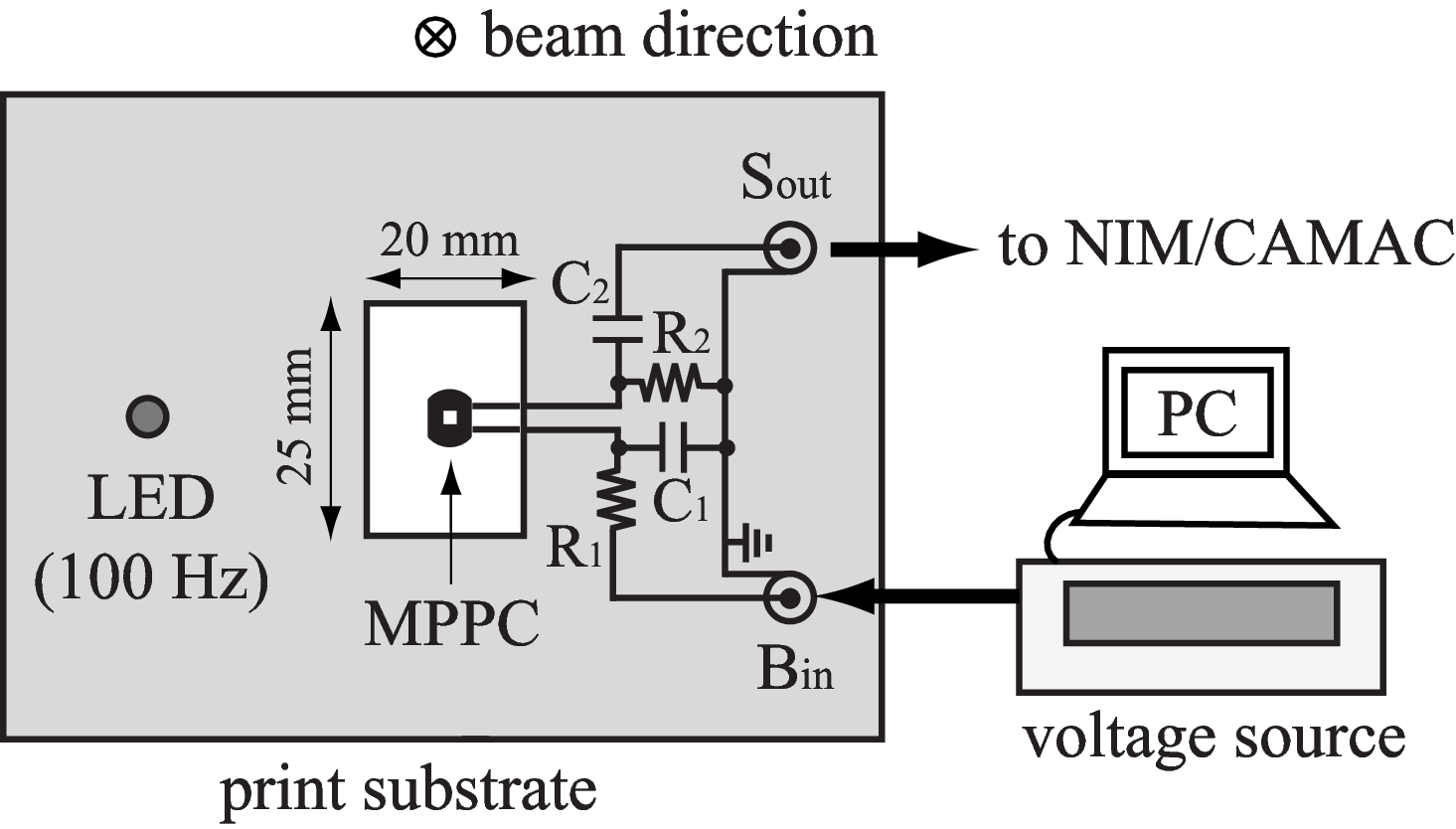}
  \end{center}
  \caption{Illustration of the print substrate 
    prepared for the experiment. 
    A proton beam hit the MPPC sample 
    mounted on the substrate from the direction 
    perpendicular to the entrance face of the sample.
    The substrate has a $20\times25$~mm$^2$ hole around the sample 
    to avoid radio-activation. 
    A voltage source was connected to the 
    connector B$_{\rm in}$ to apply 
    a negative reverse voltage to the sample. 
    Signals from the sample were read out 
    from the connector S$_{\rm out}$. 
    Capacitances (C$_1$:0.047~$\mu$F,   C$_2$:0.1~$\mu$F) and 
    resistors (R$_1$:10~k$\Omega$, R$_2$:1~k$\Omega$) 
    were used for noise-filter circuits. A blue LED mounted on 
    the substrate was employed as a light source 
    for measurement of pulse-height distributions.}
  \label{fig:substrate}
\end{figure}

One of the MPPC samples was mounted on a print substrate 
as illustrated in Fig.~\ref{fig:substrate}. 
Protons were directed onto the sample 
from the direction perpendicular to the entrance face of the sample. 
To avoid radio-activation of materials other than MPPC 
caused by proton irradiation, 
a 20$\times$25~mm$^2$ hole was made in the substrate.
A blue LED (NICHIA NSPB320BS, $\lambda = 470$~nm) on the substrate was 
used as a light source for measurements of pulse-height distributions.
A voltage source, KEITHLEY model 2400, was employed 
in order to apply a negative reverse voltage to the MPPC sample.  
The leakage current of the sample was 
measured with the voltage source and  
recorded by the laptop computer connected to 
the voltage source via GP-IB communications.  
Signals from the MPPC sample were sent to a readout circuit, 
consisting of NIM and CAMAC modules,  
to measure pulse-height spectra and dark-noise rates. 
The resistors and capacitors on the substrate were used 
as noise-filter circuits. 

\subsection{Proton beam}

The experiment was carried out at the Research Center for Nuclear Physics, 
Osaka University by using the 53.3~MeV proton beam from the AVF cyclotron. 
Protons were extracted from a vacuum beam pipe of the H-course beamline 
through a thin aluminum window to the experimental area. 
The beam size was set to be a rectangular shape of 6$\times$8~mm$^2$ 
by adjusting the several beamline slits, 
which was enough to cover the active area of MPPC, 1$\times$1~mm$^2$. 
The beam current was first tuned to be 2~nA at the beam stopper 
placed downstream of the final slit. 
Then the beam intensity was further reduced and controlled 
by inserting several mesh-type attenuators in the beamline 
located between the ion source and AVF cyclotron, 
so that the proper proton beam flux was obtained 
for the irradiation experiment (10$^4$-10$^5$~p/mm/s). 
The uniformity of the beam intensity near the beam center  
was measured with a copper collimator and plastic scintillators.
From this measurement, we confirmed that the position dependence 
of the beam intensity varied by less than 5\% 
in the region of 4$\times$4~mm$^2$ around the beam center.

\subsection{Experimental setup}

\begin{figure*}[t]
  \begin{center}
    \includegraphics[width=10cm]{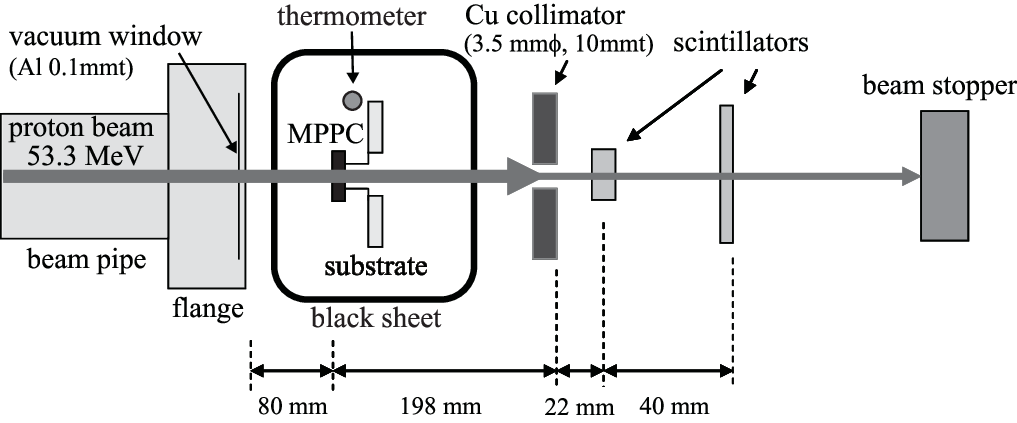}
  \end{center}
  \caption{Experimental setup for the proton irradiation. 
    The proton beam passed from left to right in this figure. 
    The MPPC sample mounted on the substrate was placed 
    downstream of the beam-pipe flange. 
    A platinum thermometer was fixed near the sample 
    for monitoring temperatures during the experiment. 
    Two plastic scintillators placed behind the copper collimator 
    were employed to monitor the beam intensity (see text). }
  \label{fig:setup}
\end{figure*}

Fig.~\ref{fig:setup} shows the experimental setup for this proton irradiation. 
A MPPC sample mounted on the substrate was placed just after 
the vacuum window of the beam pipe.  
The sample was covered with a black sheet (200~$\mu$m thickness) 
for light shielding. 
We set a platinum thermometer near the MPPC sample 
in order to monitor variations of the temperature 
inside of the black sheet during the experiment. 
The proton beam emerging from the beam pipe passed through the MPPC sample, 
and then it was trimmed with a copper collimator to be 3.5~mm$\phi$. 
Two plastic scintillators placed downstream of the collimator 
were used to monitor the number of protons 
contained in the collimated beam. 
The proton beam was dumped in the beam stopper 
placed at the end of the beamline.

The intensity of the beam impinging on the active area of 
the MPPC sample was estimated from 
the total counts of the coincidence signals of 
the plastic scintillators and the measurement times.
Since the beam intensity was almost flat 
in the region of 4$\times$4~mm$^2$ around the beam center,  
we obtained the intensity from the coincidence rate 
by taking the ratio between the active area of the MPPC sample and 
the area of the collimator hole, $R_A=0.1040\pm0.0015$, into account, 
where we assumed the machining accuracy of the collimator hole to be 50~$\mu$m.  
The following corrections were also taken into account:
beam loss due to the scattering at the ceramic package of 
the MPPC sample and the black sheet ($C_s$), 
and the beam divergence induced difference of the beam density 
at the sample position and at the collimator position ($C_d$). 
The factor $C_s$ was obtained by the coincidence rate change 
between with and without the MPPC sample wrapped by the black sheet. 
$C_d$ was also determined by comparing the coincidence rates 
at different collimator positions.    
We estimated $C_s$ and $C_d$ to be 2.71 and 1.18, respectively. 
The considered uncertainties of the product of $C_s \cdot C_d$ are  
the coincidence rate fluctuation during the irradiation (3.4~\%) 
and accuracy of the detector and collimator positioning (0.9~\%).  
As the result, we obtained the correction factor, 
$R_A\cdot C_s\cdot C_d = 0.333\pm0.013$, 
to multiply to the coincidence rates. 
The proton beam-flux during irradiation was monitored in this way.

\begin{figure}[t]
  \begin{center}
    \includegraphics[width=7cm]{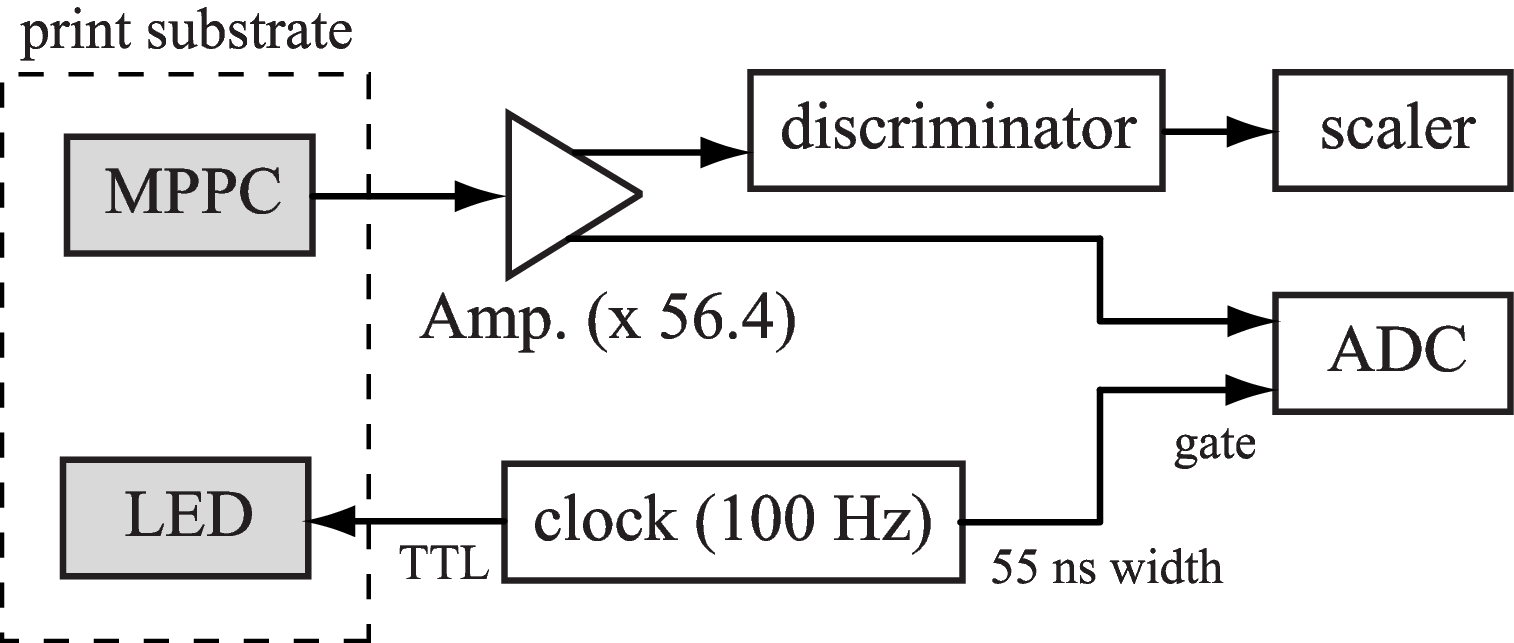}
  \end{center}
  \caption{Readout circuit for measurement of pulse height distributions 
    and dark noise. The LED was driven with a 100 Hz clock generator.
    To digitize the signals from the MPPC on the substrate, 
    we used a CAMAC ADC, 
    which is connected to the amplifier with a gain of 56.4.
    A discriminator and a scaler were used 
    for the measurement of dark-noise rates.}
  \label{fig:circuit}
\end{figure}

A readout circuit consisting of NIM and CAMAC modules 
was prepared for the measurements of pulse-height distributions 
and dark-noise rates of the MPPCs 
as illustrated in Fig.~\ref{fig:circuit}.
The blue LED was driven with TTL signals coming from 
a 100~Hz clock generator. 
Signals from the MPPC sample 
were sent to an amplifier (Phillips Model 777) 
where the gain was set to be 56.4. 
Then, these signals were integrated with a 12-bit charge-sensitive ADC 
of a CAMAC module (REPIC RPC-022).
Gate signals from the clock generator were input to the ADC module. 
The gate width, 55~ns, was determined so that entire pulse signals  
coming from the MPPC were integrated, 
where the typical fall time of the signals was 40~ns. 
The intensity of the LED light was adjusted in such a way that 
the average number of photo-electrons ($\mu$) was about 1.3. 
This value can be calculated from 
the null probability of the Poisson statistics $P_0$ $(=e^{-\mu}$), 
which was obtained from the ratio of the 
number of pedestal events to the total events 
in an ADC distribution. 
A discriminator (Phillips Model 705) and a visual scaler (Kaizu Works KN1860) 
were used for the measurement of dark-noise rates.

\begin{table*}
\begin{center}
\caption{Summary of proton irradiation for Sample~\#1 and Sample~\#2. 
The symbols represent as follows: $\phi_{\rm p}$ (flux of the proton beam), 
$t$ (irradiation time), $\Phi_{\rm p}$ (total fluence of the proton beam),  
$\Phi_{\rm eq}$ (1 MeV neutron equivalent fluence), $D$
(accumulated absorption dose), and $T$ (temperature during each measurement)
.} 
\label{tab:flux}

Sample~\#1 \\
\begin{tabular}{c||cc|ccc|c} \hline
irradiation & $\phi_{\rm p}$ (mm$^{-2}s^{-1}$)  & $t$ (min.) &  $\Phi_{\rm p}$  (mm$^{-2}$)& 
 $\Phi_{\rm eq}$ (mm$^{-2}$)  & $D$ (Gy) & $T$ ($^\circ$C) \\ \hline
before & --- & --- & --- & --- & --- & 27.6$\pm$0.1 \\
1st & $2.3\times 10^5$ & 10 & $1.4\times 10^8$ & $2.4\times 10^8$ & 21 & 28.0$\pm$0.1 \\
2nd & $2.4\times 10^5$ & 10 & $2.8\times 10^8$ & $4.8\times 10^8$ & 42 & 28.0$\pm$0.1 \\
\hline 
\end{tabular}

Sample~\#2 \\ 

\begin{tabular}{c||cc|ccc|c} \hline
irradiation & $\phi_{\rm p}$ (mm$^{-2}s^{-1}$)  & $t$ (min.) &  $\Phi_{\rm p}$  (mm$^{-2}$)& 
 $\Phi_{\rm eq}$ (mm$^{-2}$)  & $D$ (Gy) & $T$ ($^\circ$C) \\ \hline
before & --- & --- & --- & --- & --- & 27.6$\pm$0.1 \\
1st & $3.1\times 10^4$ & 10 & $1.9\times 10^7$ & $3.1\times 10^7$ & 2.8 & 27.4$\pm$0.1 \\
2nd & $3.0\times 10^4$ & 10 & $3.7\times 10^7$ & $6.2\times 10^7$ & 5.5 & 27.2$\pm$0.1 \\
3rd & $2.8\times 10^4$ & 10 & $5.3\times 10^7$ & $9.1\times 10^7$ & 8.0 & 27.0$\pm$0.1 \\
\hline 

\end{tabular}
\end{center}
\end{table*}


\subsection{Experimental procedure}
First, we checked basic performance 
of the MPPC samples, such as gain, 
current-voltage ($I$-$V$) curve 
and noise rate, before the proton irradiation. 
This performance is to be compared with that after the irradiation. 
For the next step, one of the MPPC samples 
with normal operating voltage 
was irradiated with the proton beam.
After the irradiation, 
in order to check recovery effects on radiation damage, 
the leakage current was monitored for about 1 hour  
under the condition without the beam.  
Then we measured the basic performances after the irradiation. 
This process was repeated several times.

We used two different beam intensities for the irradiation.
One of the MPPC samples, Sample~\#1, was irradiated 
with a proton beam flux ($\phi_{\rm p}$) of
$2.3\times 10^5$~p/mm$^2$/s. 
The second MPPC, Sample~\#2, was irradiated with a lower flux, 
$\phi_{\rm p}=3.0\times 10^4$~p/mm$^2$/s.  
One irradiation took 10 minutes for both the samples. 
For Sample~\#1, we repeated the proton irradiation twice, 
thereby estimated the proton fluence ($\Phi_{\rm p}$) is 
$2.8\times 10^8$~p/mm$^2$ in total.
On the other hand, three times irradiation were made for Sample~\#2, 
thus the total fluence is $5.3\times 10^7$~p/mm$^2$.
The beam flux and total fluence in each irradiation 
are summarized in Table \ref{tab:flux}.

Instead of the proton fluence $\Phi_{\rm p}$, 
it may be useful to represent a total fluence with  
1 MeV neutron equivalent fluence ($\Phi_{\rm eq}$). 
This quantity is commonly used in order to evaluate 
damage levels caused by different radiation sources 
and different energies based on the NIEL scaling hypothesis, 
which is the assumption that damage level in silicon bulk is related to 
the cross-section of processes 
with Non-Ionization Energy Loss~\cite{Lindstrom:2002gb}.
The $\Phi_{\rm eq}$ can be obtained with a hardness factor $\kappa$  
as $\Phi_{\rm eq}=\kappa\cdot\Phi_{\rm p}$,  
where the value of the $\kappa$ for 53.3 MeV protons is about 1.7 
obtained from the displacement damage function described 
in the reference~\cite{Lindstrom:2002gb}.
Hence, we estimated the $\Phi_{\rm eq}$ for Sample~\#1 to be 
$4.8\times 10^8$~n/mm$^2$, and 
$9.1\times 10^7$~n/mm$^2$ for Sample~\#2, respectively.

The absorption dose ($D$) due to the proton irradiation 
can be estimated from the proton fluence $\Phi_{\rm p}$ by taking 
the mass stopping power of 53.3~MeV protons in silicon, 
$9.38$ MeV$\cdot$cm$^2$/g, into account. 
The estimated dose rates for both Sample~\#1 and Sample~\#2  
were 130~Gy/h and 16~Gy/h, respectively. Thus, 
the total radiation dose was 42~Gy for Sample~\#1, and 8.0~Gy for 
Sample~\#2.

The temperature in the experimental area 
was controlled to be stable. 
During the experiment 
the temperature near the MPPC samples ($T$) was 
in the range 27.0 - 28.0~$^\circ$C. 
Note that we estimated the maximum temperature increment 
caused by the proton beam in the MPPC volume to be 0.03~$^\circ$C 
in the case of 10 minutes irradiation, which was calculated from 
the beam energy deposition in the MPPC and the specific heats 
of the device materials. Thus, we concluded that the temperature increase 
due to the irradiation was negligible compared with 
the temperature fluctuation in the experimental area.
The temperatures in each measurement are 
also summarized in Table \ref{tab:flux}.


\section{Result}

\subsection{Variations of the leakage current} \label{sec:leakcurrent}

\begin{figure*}[t]
  \begin{center}
    \includegraphics[width=14cm]{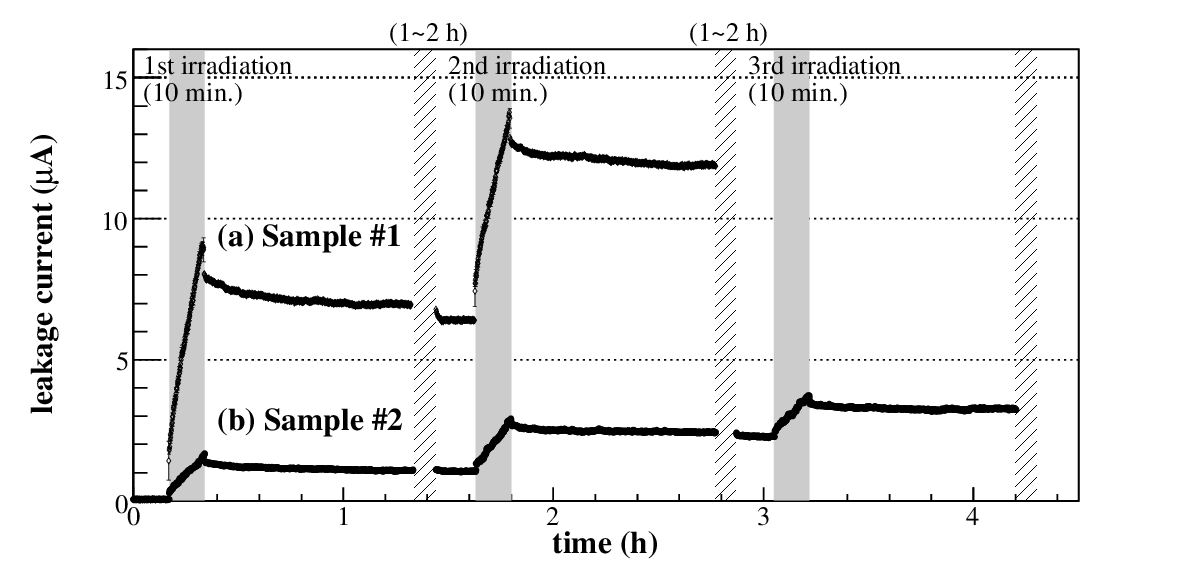}
  \end{center}
  \caption{Variations of the leakage current for 
    (a) Sample~\#1 and (b) Sample~\#2. 
    The fixed operating voltages, 
    69.75~V for Sample~\#1 and 69.58~V for Sample \#2, 
    were applied to the samples with the voltage source.
    The proton irradiation was performed  
    during the times represented as the hatched regions. 
    In the times shown as the shaded regions, 
    no data was recorded since measurements 
    for gains, current-voltage ($I$-$V$) curves 
    and noise rates, were carried out, taking typically 1-2 hours. 
    Note that the actual voltages applied to the samples 
    changed during the measurement because of a voltage drop 
    at the resistors in the filter circuit (see text).
  }
  \label{fig:IT}
\end{figure*}

Fig.~\ref{fig:IT}(a) shows the variations of the leakage current of 
Sample~\#1 during the experiment. 
The measurement of the leakage current 
was started 10 minutes before the first irradiation,  
where the leakage current was 
about $0.05$~$\mu$A at the operating voltage.  
We irradiated the sample with the proton beam for 10 minutes 
(see the leftmost hatched region in the figure). 
During the first irradiation, 
the current almost linearly increased with time due to radiation damage. 
Step-like changes at the beginning and ending of the irradiation, 
$\sim$1.0~$\mu$A, 
would be caused by free carriers 
generated due to the ionizing processes of the beam protons 
traversing in silicon. 
After the beam stops, the leakage current gradually 
decreased with time. 
This indicates recovery phenomenon 
from the radiation damage, 
although the recovery was not completed 
to the original condition within hours. 
The second irradiation was performed after the measurements of 
pulse-height distributions, $I$-$V$ curves and dark noise.  
Again the leakage current increased linearly with time,  
and a similar recovery effect 
is also seen in the second irradiation.  
Note that we applied the fixed operating voltage 
with the voltage source during the measurement; 
thus, the increase of the leakage current due to the 
radiation damage caused a gain reduction 
due to a voltage drop at the resistors $R_1$ and $R_2$ 
in the filter circuit (see Fig.~\ref{fig:substrate}).
Such effect should be corrected for performance studies  
and this will be discussed later. 
The result of leakage-current variations for Sample~\#2, 
which was irradiated with a lower beam flux, 
is shown in Fig.~\ref{fig:IT}(b).
Although the increasing rate during the irradiation was low 
compared to that for the Sample~\#1 irradiation 
because of the lower beam flux, 
we have found a similar tendency as in the Sample~\#1 irradiation,  
that is, the linear increase of the leakage current and 
the recovery effect. 
The offset of the leakage current due to 
the ionizing processes of protons in silicon
was about 0.3~$\mu$A in this case. 

As we mentioned, 
the effect of the voltage drop at the resistors $R_1$ and $R_2$  
should be corrected.  
Since the leakage current ($I_{\rm leak}$) 
flows the resistors arranged in series, 
the voltage drop is expressed as $I_{\rm leak}(R_1+R_2)$. 
Hence, actual reverse voltages of the MPPC ($V_{\rm MPPC}$) 
is given by 
\begin{equation}
V_{\rm MPPC} = V_{\rm source} - I_{\rm leak}(R_1+R_2), 
\end{equation}
where $V_{\rm source}$ denotes the voltage applied 
by the voltage source. 
In the following part of this paper, 
we will express reverse voltages as $V_{\rm MPPC}$ 
in place of $V_{\rm source}$.

\begin{figure}[t]
  \begin{center}
    \includegraphics[width=7cm]{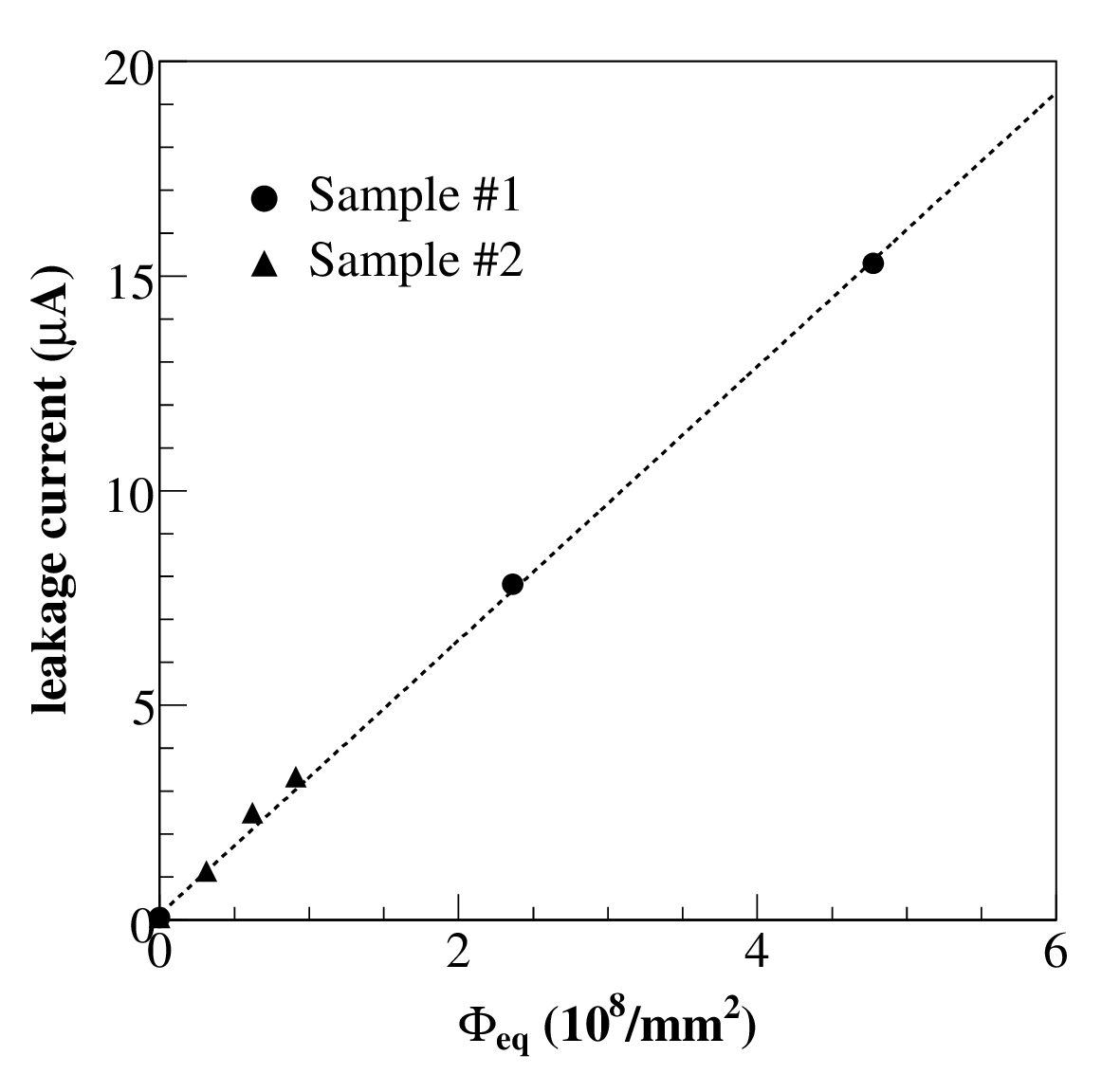}
  \end{center}
  \caption{Plots of the leakage current under the operating voltage 
    measured for 1 hour after irradiation 
    as a function of 1 MeV neutron equivalent fluence, $\Phi_{\rm eq}$,  
    for Sample~\#1 (filled circles) and Sample~\#2 (filled triangles).
    The straight line shows the function fitted to the data for
    Sample~\#1.
  }
  \label{fig:I-Flux}
\end{figure}

Fig.~\ref{fig:I-Flux} shows 
the leakage current of both samples 
under the operating voltages ($V_{\rm MPPC}=69.75~{\rm V}$ 
for Sample~\#1 and $V_{\rm MPPC}=69.58~{\rm V}$ for Sample~\#2)
as a function of the 1 MeV neutron equivalent fluence $\Phi_{\rm eq}$,  
where the data has been taken 
after 1 hour from each end of irradiation. 
From these plots we have found that 
the leakage current increases linearly with the fluence. 
The rates of increase for the both samples 
are almost the same although the beam flux for Sample \#1 
is about eight times higher than for Sample \#2. 
The straight line in the figure 
shows the linear function 
fitted to the data of Sample~\#1, which gives us the relation between 
the leakage current and the irradiation dose,  
\begin{equation}
I_{\rm leak}[\mu{\rm A}] = 3.2\times10^{-8}\cdot\Phi_{\rm eq}[{\rm n/mm^2}].
\end{equation}
This information would be useful to compare radiation hardness of 
other PPD devices with that of MPPC.

\begin{figure}[t]
  \begin{center}
    \includegraphics[width=7cm]{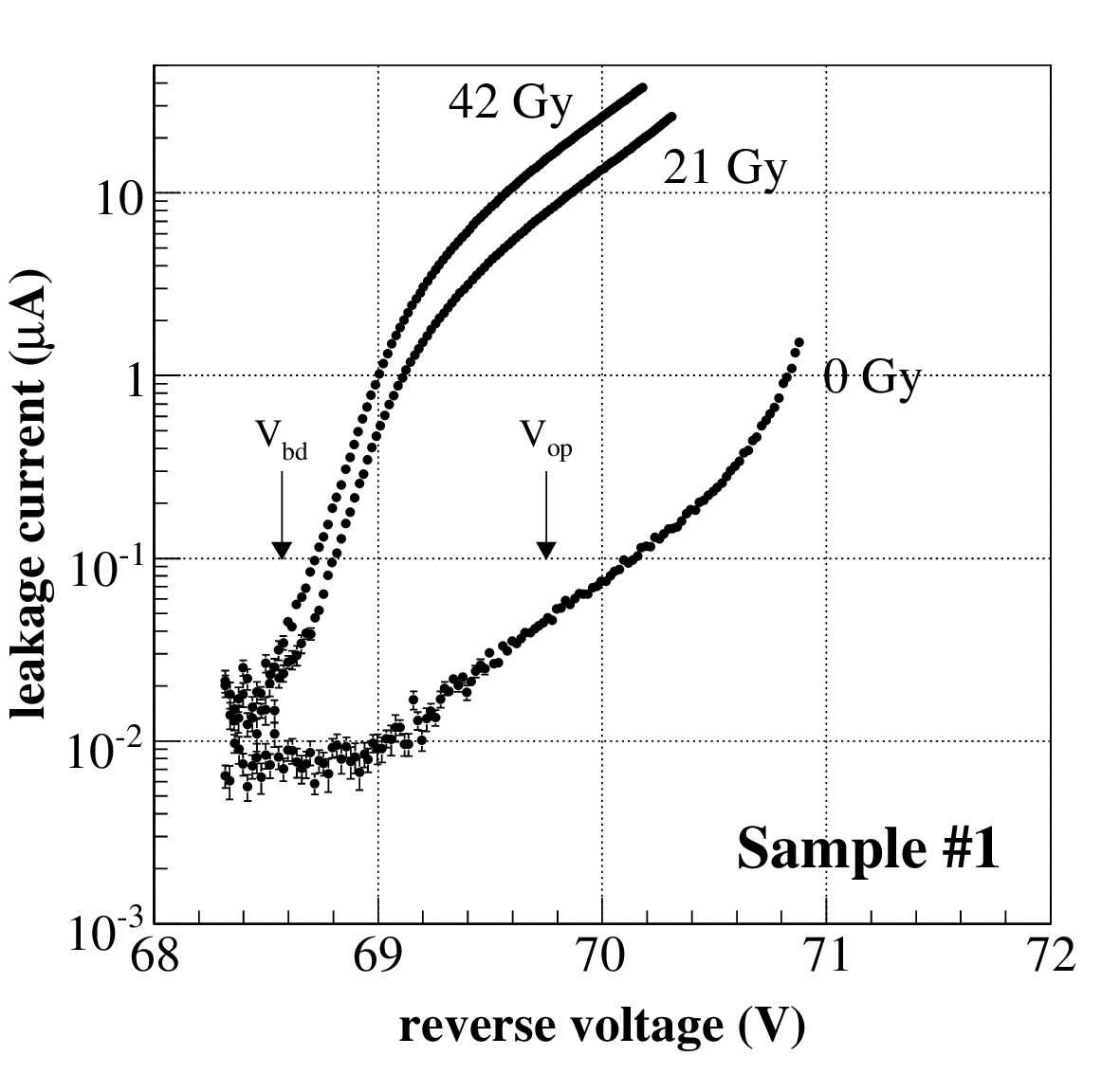}
    \includegraphics[width=7cm]{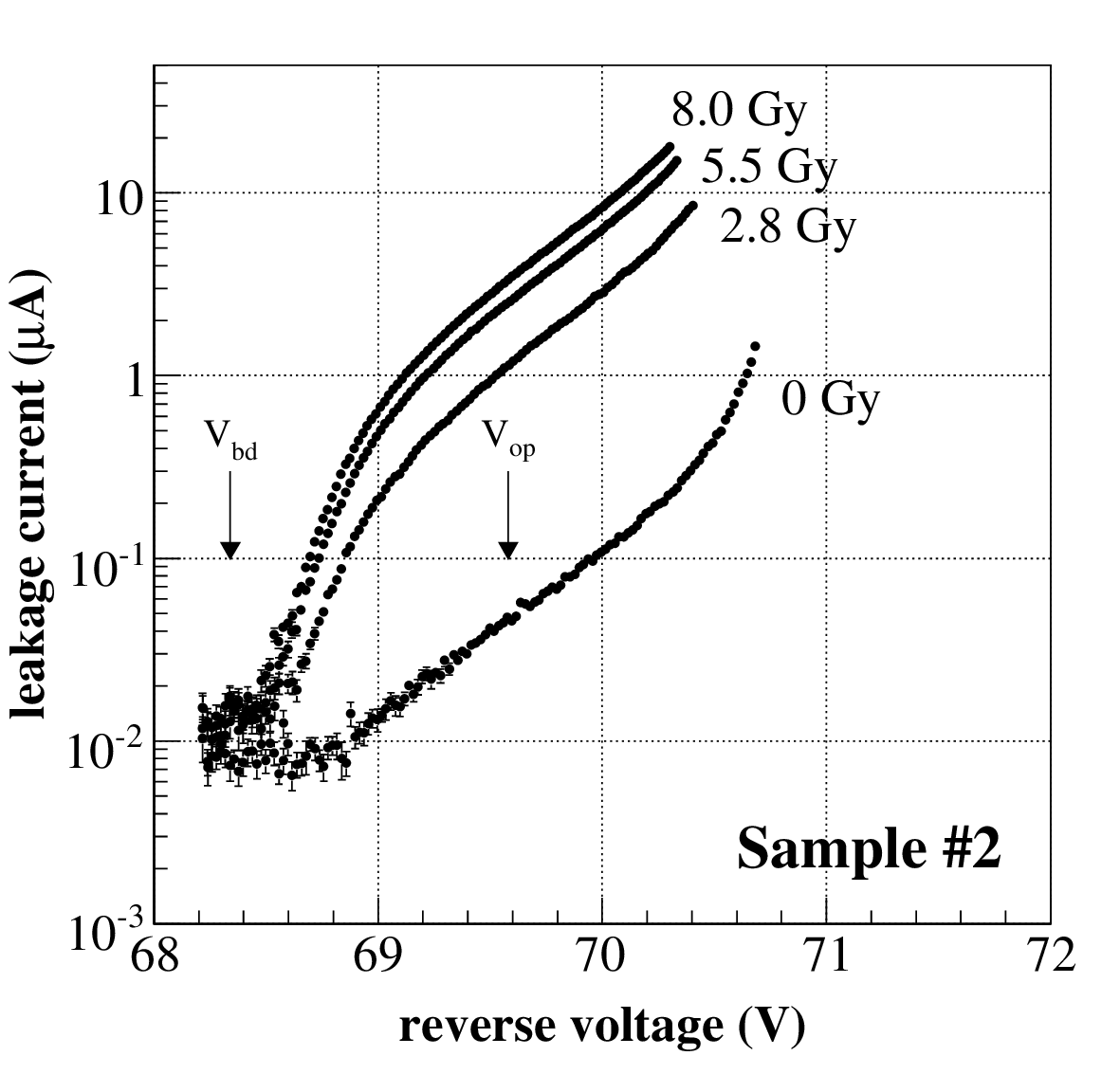}
  \end{center}
  \caption{$I$-$V$ curves of Sample~\#1 (top) and Sample~\#2 (bottom) 
    for different radiation doses. The symbol $V_{\rm bd}$ denotes 
    the breakdown voltage of each sample 
    measured before the irradiation at 27.6 $^\circ$C. 
    The symbol $V_{\rm op}$ indicates 
    the operating voltage given by the manufacturer. 
  }
  \label{fig:IV}
\end{figure}

The $I$-$V$ curves were 
measured before and after the irradiation without the LED light. 
The results are shown in Fig.~\ref{fig:IV}. 
The symbols $V_{\rm bd}$ in the figure  
denote the breakdown voltages measured before the irradiation; 
they are 68.57$\pm$0.01 V for Sample~\#1 and 
68.34$\pm$0.01~V for Sample~\#2 at 27.6 $^\circ$C. 
From these plots, the leakage currents have rapidly 
increased with reverse voltages after the irradiation  
in comparison with that before the irradiation. 
This tendency is more significant for higher doses.


\subsection{Photon-counting capability} \label{subsec:PCC}

\begin{figure*}[t]
  \begin{center}
    \includegraphics[width=11cm]{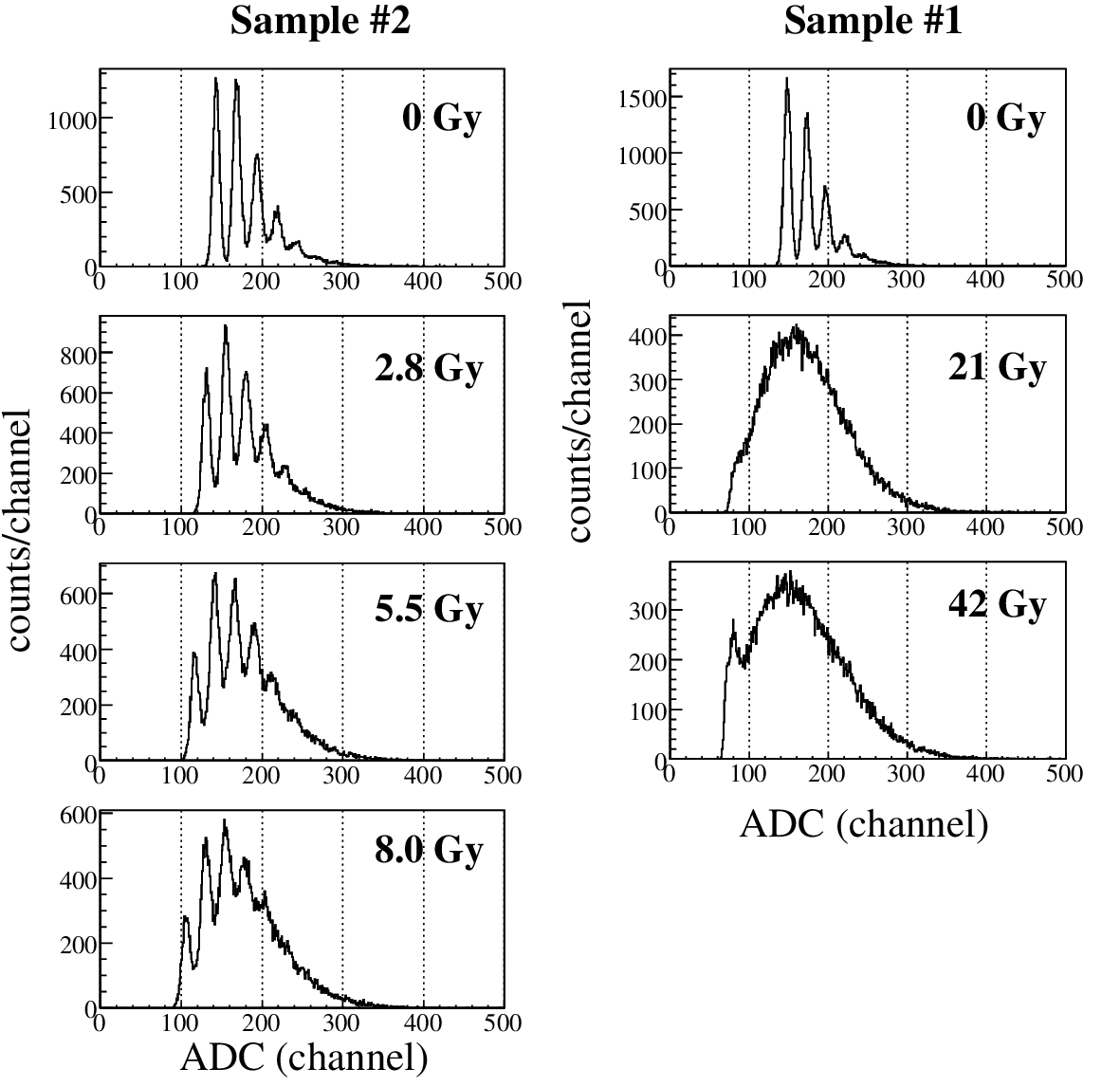}
  \caption{
    Pulse height distributions measured 
    with a blue LED for different irradiated doses. 
    The operating voltages were applied to both of the samples.
    The position of peaks located at the lowest level 
    in each histogram, except for the 21~Gy and 42~Gy spectra,
    correspond to the pedestals.
 }
  \label{fig:ADC_dist}
  \end{center}
\end{figure*}

Next we discuss the effects on photon-counting capability, 
which is one of the features of MPPCs. 
The pulse-height distributions of the irradiated samples 
were measured by flushing the blue LED. 
Fig.~\ref{fig:ADC_dist} shows the distributions 
for different irradiated doses.  
Before the irradiation the distributions for both samples 
show the structure with clear discrete peaks, which includes    
a pedestal peak, a single photon peak and so on, 
and those peaks are well separated each other. 
In contrast, the situation changes after the irradiation. 
First, the position of the pedestal peak shifts slightly 
toward lower channels because of the baseline shift 
coming from high dark counting rates. 
Note that the dark counting rate of MPPC  
increases after the irradiation, which is the main reason for 
the increase of the leakage current. 
This is because production of the lattice defects due to radiation damage 
gives higher probability of thermal carrier generation and 
delayed pulse coming from trapped carriers 
in the defects (afterpulsing). 
Actually, the measurement of the counting rates without the LED 
showed that the counting rates drastically increased from 
270~kHz to 6.8~MHz after the 2.8~Gy irradiation  
(Although we could not measure the counting rates 
for higher doses due to limitation of the scaler performance, 
the counting rates would be expected to be more than 10~MHz).
Another thing is that the distributions are contaminated with 
accidental backgrounds coming from dark noise and afterpulsing 
as the dose increases.   
This effect smears the separation of the peaks 
in the pulse-height distributions as seen in Fig.~\ref{fig:ADC_dist}  
since some of noise pulses are not integrated over the entire charge. 
Gain uniformity among the individual APD pixels 
may also be deteriorated by radiation damage 
because if the extent of the damage differs among the pixels, 
average leakage current through each quenching resistor varies. 
This results in gain variations among the pixels 
because of the different voltage drops at the quenching resistors, 
and thus the peaks could be broadened. For these reasons, 
the peak structure has completely disappeared 
for the distributions of more than 21~Gy 
in the case of 55~ns gate width. 
This means that the capability of the photon-counting 
by measuring pulse-height spectra was lost due to the radiation damage.

\subsection{Gain}

The gain variations were evaluated before and after the irradiation. 
We define the gain of the MPPCs as the ratio of 
the collected charge ($Q$) and the electron charge ($e$), $Q/e$. 
The reason is that a single carrier generated by an incident photon 
is multiplied due to Geiger discharge, and then  
the collected charge in a single pixel is extracted as a output signal.  
In the data analysis the charge $Q$ was estimated   
from the gain of the amplifier, 56.4, 
and the charge corresponding to single-photon signals. 
The charge of single-photon signals was obtained by the fit 
to the pulse-height distributions with a multi-Gaussian function.

\begin{figure}[t]
  \begin{center}
    \includegraphics[width=7.5cm]{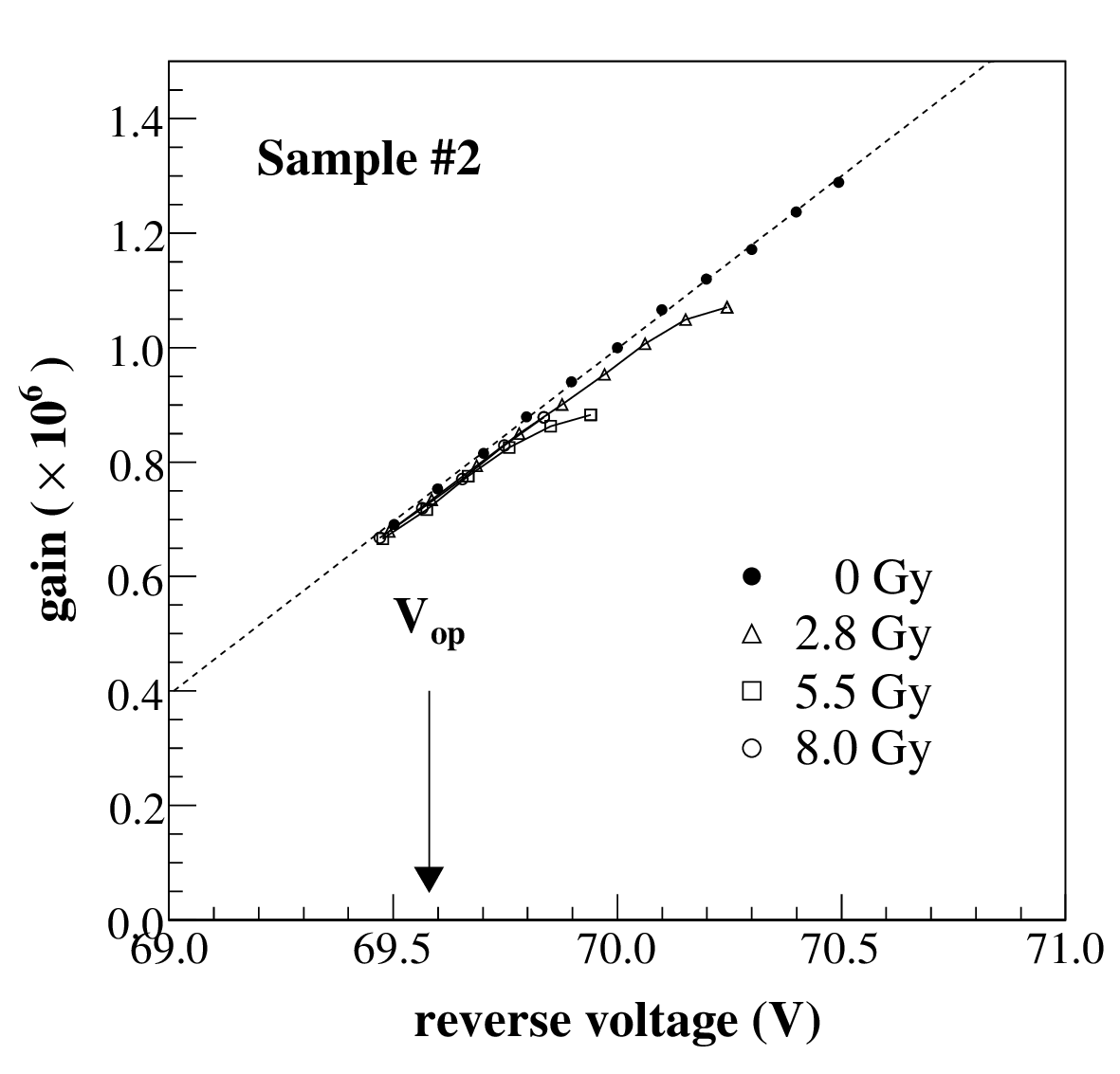}
    \caption{Gains of Sample~\#2 as a function of reverse voltage for different
      radiation doses. 
      Dotted line shows the function~(\ref{eq:gain}) fitted to the data 
      before the irradiation.
      The symbol $V_{\rm op}$ denotes the operating voltage.
  }
  \label{fig:Gain-V}
  \end{center}
\end{figure}

Fig.~\ref{fig:Gain-V} shows the gains as a function of reverse voltages 
for different radiation doses. 
Note that only the gains for Sample~\#2 was obtained 
since we observed no peak structure 
in the distributions of Sample~\#1 after the irradiation.
The gain before the irradiation shows 
a linear dependence with the reverse voltage ($V$)
since the gain of a single pixel ($G$) 
increases with $V$ as 
\begin{equation}
G=Q/e=C(V-V_{\rm bd})/e \label{eq:gain}
\end{equation}
where $C$ denotes the average capacitance of a single pixel, and $V_{\rm bd}$ 
is the breakdown voltage. 
The capacitance $C$ was estimated to be 0.096~pF 
from the result of a linear fitting, which is shown as 
the dotted line in Fig.~\ref{fig:Gain-V}.
After the irradiation, the data points  
slightly deviate from a linear relation at higher voltages, 
where the simple fitting with a multi-Gaussian function 
may be inadequate because of significant accidental backgrounds 
coming from dark noise and afterpulsing. 
However, at the operating voltage where those backgrounds would be less affected, 
gain change is not significant (within 3\% level). 
The result shows that 
the effect to the gain due to radiation damage 
is small at least up to 8.0~Gy irradiation. 
Note that there is no data point at higher voltages after the irradiation  
because of the disappearance of the peak structure in the pulse-height distributions.

\subsection{Pulse-height resolution after more than a year from the irradiation}

\begin{figure}[t]
  \begin{center}
\includegraphics[width=5cm]{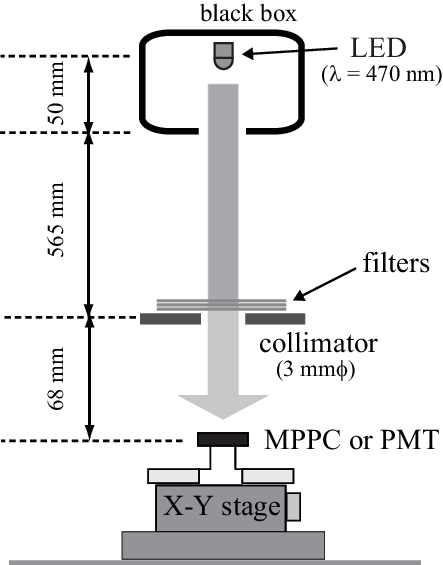}
  \caption{Setup for the measurement of the pulse-height resolution. 
    Collimated LED light was illuminated one of the MPPC samples, 
    which was mounted on a movable X-Y stage. 
    The filters on the collimator were used to control the light intensity. 
    Absolute number of photons entering onto the active area of the samples 
    was obtained with a photo-multiplier tube (PMT) prior to the measurement. 
    Note that the size is not to scale.
  }
  \label{fig:setup_PDE}
  \end{center}
\end{figure}

The pulse-height resolution of photo-sensors 
is one of the most important properties 
in the case of measurements for the incident-particle's energy. 
In order to check whether the radiation damage could 
affect to the pulse-height resolution of the MPPCs, 
we measured the pulse-height distributions 
of the irradiated samples 
by illuminating with a LED light 
($\sim$ 200~photons/mm$^2$).  
Spectral changes due to the damage 
can be found by comparing to the distribution 
taken with a non-irradiated reference sample.  
This measurement has been performed 
after 430~days from the irradiation experiment.  
The irradiated samples, 
which have been kept in a desiccator at room temperature, 
have partially recovered from the radiation damage 
because the leakage current at the operating voltage  
decreased from 12.0~$\mu$A to 7.1~$\mu$A for Sample~\#1, 
and from 3.3~$\mu$A to 1.6~$\mu$A for Sample~\#2, respectively.

Fig.~\ref{fig:setup_PDE} shows the setup for the measurement. 
The entire setup was placed in a thermostat oven to ensure 
temperature stability (25~$^\circ$C) and light-tightness. 
The LED light was collimated with a collimator having the hole size of 3 mm$\phi$ 
and illuminated one of the MPPC samples. 
We aligned the sample with the center of the light spot 
by using a X-Y stage moving with an accuracy of 5 $\mu$m. 
The intensity of incident light was controlled with 
filters placed on the collimator.

Prior to the measurement, 
we estimated the absolute number of the photons 
entering to the active area of the MPPC ($\rm N_{\rm ph}$)  
with a green-extended photo-multiplier tube (PMT), 
Hamamatsu H3178-61. 
The entrance window of the PMT was masked with black slits 
having the area of 1.0$\times$1.0~mm$^2$  
so as to meet the requirement for the same active area of the MPPC.
The quantum efficiency of the PMT at the wavelength of 470~nm, 
which corresponds to the peak wavelength of the LED,  
was measured to be 26.0~\% by the manufacturer.
Two different intensities of light, $\rm N_{\rm ph}=11.2\pm0.2$~photons/mm$^{2}$ 
and $\rm N_{\rm ph}=193\pm4$~photons/mm$^{2}$,  
were illuminated in the measurement.

\begin{figure*}[bt]
  \begin{center}
    \includegraphics[width=12cm]{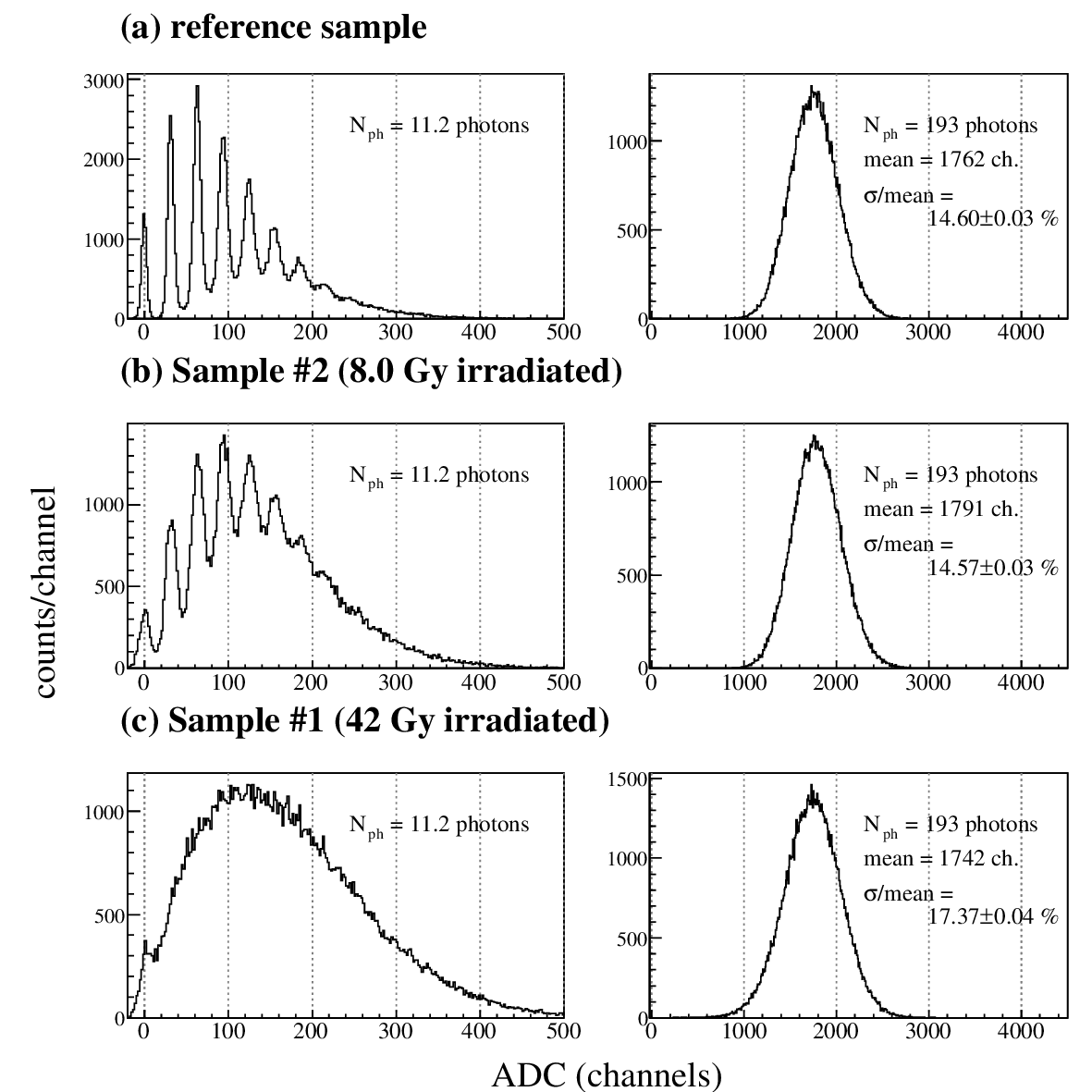}
  \caption{Pedestal subtracted ADC distributions 
    for the reference sample (a), Sample~\#2 (b) and Sample~\#1 (c).   
    Histograms in the left column show the distributions 
    measured with low-intensity light ($11.2\pm0.2$~photons/mm$^2$),  
    and in the right column the distributions 
    obtained with high-intensity light ($193\pm4$~photons/mm$^{2}$) are presented. 
    The mean values and the relative widths 
    of the distributions obtained by Gaussian fittings  
    are shown in the right figures, 
    where the errors of the relative widths were estimated from 
    the fittings.
  }
  \label{fig:LowHigh}
  \end{center}
\end{figure*}

The result is shown in Fig.~\ref{fig:LowHigh}, where  
pedestal-subtracted ADC distributions of the three samples 
(reference sample, Sample~\#2, and Sample~\#1) are presented 
for the different light intensities. 
In the low-intensity case,  
the peak structure in the distribution of the irradiated samples 
was smeared due to the radiation damage as discussed in the section \ref{subsec:PCC}, 
although more than a year have elapsed from the irradiation 
and the recovery effect in the leakage current was observed. 
In the high-intensity case, the distributions show a Gaussian shape 
having the mean values of about 1760 channels, 
where the repeatability of the mean-value measurement 
was evaluated to be about 2\%.  
Thus, we found that there was no significant change in the mean value 
of the pulse height-distribution among the three samples,  
suggesting that the irradiated samples work well as a photo-sensor 
even after photon-counting capability was lost. 
In contrast, the relative width of the distribution 
slightly widened for Sample~\#1 whose irradiated dose was 42~Gy in total.  
The measured widths were $14.60\pm0.03$~\% 
for the reference sample, $14.57\pm0.03$~\% for Sample~\#2, 
and $17.37\pm0.04$~\% for Sample~\#1, respectively, 
where the errors were estimated from Gaussian fittings. 
Assuming that the fluctuation caused by 
radiation damage contributed to the width in quadratic sum, 
the contribution was estimated to be about 9~\%, where 
the effect of the broadened baseline fluctuation was 
negligible based on a comparison of the pedestal widths 
of the reference sample and Sample \#1. 
In principal, the relative width, which arises because of  
fluctuations in the number of fired APD-pixels, 
is given by Poisson statistics. 
Thus, the difference of the relative width 
would be not due to the variability of individual devices, 
but given due to the radiation damage.  
Some details about the pulse-height resolution 
will be discussed later.


\section{Discussions}

The leakage current of the MPPC 
increased with total irradiated dose as shown in Fig.~\ref{fig:I-Flux}. 
This behavior can be interpreted as an increase of thermal excitations 
of carriers via intermediate states in the forbidden gap 
created due to radiation damage. 
Such effect is known to be categorized into 
bulk damage and surface damage, 
where the bulk damage is caused by defects in the Si bulk and 
the surface damage is created at the Si-SiO$_2$ interface. 
In the case of proton irradiation,  
both of them can contribute to the increase of the leakage current. 
In order to understand the mechanism of proton-induced damage, 
studies of neutron irradiation would be helpful 
since the surface damage does not contribute
to neutron-induced radiation damage~\cite{Moll}.

Photon-counting capability, which is one of 
the features of MPPCs, was gradually deteriorated 
with the irradiation dose and completely 
lost after 21~Gy irradiation  
due to baseline fluctuations and noise pileup 
as a result of increasing dark counting rates. 
This effect would be problematic for some applications 
in high-energy physics experiments. 
For example, when one uses the MPPC 
as a photon-counting device 
for a ring-imaging Cherenkov detector (RICH), 
which detects weak light generated from Aerogel radiators, 
the reconstruction of ring images becomes more difficult due to the 
increase of fake hits. 
This point should be carefully considered. 
On the other hand, for the detectors which is expected 
to yield more light ($\ge 10$ photo-electrons),  
such as scintillators with wavelength-shifting (WLS) fiber readout, 
the damage effect might be acceptable at least up to 8.0~Gy 
because no significant change was observed in the gain and 
the pulse-height distribution.

We have found an indication of 
the deterioration of the pulse-height resolution 
for the sample irradiated with 42~Gy. 
Although we have not understood the mechanism of 
the deterioration,  
there are two possible explanations: 
(1) change of gain uniformity among the individual pixels and 
(2) increase of afterpulsing.  
The afterpulsing effect 
would be expected to increase after the irradiation because of radiation damage. 
Hence, increase of afterpulsing could affect 
to the pulse height distributions. 
The deterioration of the pulse-height resolution 
is an issue to be concerned 
in the case of energy-measurement applications.

\section{Summary}

The effects on MPPCs caused by proton irradiation 
has been discussed from a view point of applications 
in high radiation environment.
In order to investigate the effects of radiation damage, 
we have performed an irradiation experiment 
by using a 53.3~MeV proton beam.  
The linear increase of the leakage current 
has been observed due to radiation damage during the irradiation.  
We found recovery effects from the radiation damage, 
however it has not completed even after a year from the irradiation.
The result of the gain measurement shows 
no significant change in the gain at least up to 8.0~Gy 
(9.1$\times 10^7$~n/mm$^2$ in $\Phi_{\rm eq}$). 
From the measurement of pulse height distribution,  
we found the photon-counting capability was completely lost 
due to baseline fluctuations and noise pile-up 
after 21~Gy irradiation (2.4$\times 10^8$~n/mm$^2$ in $\Phi_{\rm eq}$).  
This effect might be problematic in the case of 
applications for low light-yield detectors such as RICH counter. 
We found that the pulse-height resolution was slightly deteriorated 
after 42~Gy irradiation (4.8$\times 10^8$~n/mm$^2$ in $\Phi_{\rm eq}$),  
where the measurement has been performed after 430 days from the irradiation experiment
with slightly intense LED light ($\sim 200$~photons/mm$^2$). 
This effect should be considered 
in the case of energy-measurement applications.

\section*{Acknowledgements}

We would like to thank the members of 
the photon sensor group of the 
KEK Detector Technology Project 
for valuable discussions on radiation damage of the MPPCs.  
We gratefully acknowledge the RCNP cyclotron staff 
for their kind cooperation in the beam preparation.
Our thanks go as well to Dr. M.~Hasinoff 
(University of British Columbia) 
who made valuable comments to the paper.  
The present work was supported in part by 
the Ministry of Education, Science, 
Sports and Culture of Japan with the Grant-in-Aid for 
Scientific Research No. 17204020.

\end{document}